\documentstyle[12pt]{article}

\begin{document}

\title{
Energy Storage in a Hamiltonian System
in Partial Contact with a Heat Bath
}

\author{
Naoko NAKAGAWA \thanks
{Department of Mathematical Sciences,
Ibaraki University,
Mito, Ibaraki 310-8512} 
\and
Kunihiko KANEKO \thanks
{\small Department of Pure and Applied Sciences,
College of Arts and Sciences,
University of Tokyo,
Tokyo 153}
}
\maketitle

\begin{abstract}
To understand the mechanism allowing for the
long-term storage of excess energy in proteins,
we study a Hamiltonian system consisting of several coupled pendula
in partial contact with a heat bath.  
It is found that energy storage is possible when
the motion of each pendulum switches between oscillatory (vibrational) 
and rotational (phase-slip) modes.   
The storage time increases almost exponentially to
the square root of the injected energy.
The relevance of our mechanism to protein motors is discussed.
\end{abstract}


Proteins are among the most important biopolymers for living systems.
They transform chemical energy to mechanical energy, and vice versa,
and contribute to biological functions.  However, the question of how
proteins work dynamically remains unanswered.
Recently, a noteworthy experiment concerning protein motors was performed
\cite{Ishijima_Yanagida}.
In this experiment, the working process of a {\it single molecule} 
was directly investigated.
The results suggest that proteins often store energy obtained 
from a reaction with ATP (adenosine triphosphate)
and use it later (e.g., for enzymatic reactions with other proteins).
The interval for energy storage was found to sometimes be very long, 
up to the order of seconds, while typical timescales for normal vibrations
are several picoseconds. 
How can proteins store excess energy for such a long time,
somehow overcoming the relaxation process
toward thermal equilibrium?
In order for a protein to store energy for a sufficiently long time, 
energy must be absorbed into a certain part of the protein,
in accordance with its own dynamics.
Furthermore, some characteristic type of dynamics is required to store the
excess energy without losing it to the surrounding aqueous solution.

As a first approach to understand the working mechanism of proteins, 
we construct a Hamiltonian system (in partial contact with 
a heat bath), which 
stores energy for a given time span, in spite of 
its eventual relaxation to thermal equilibrium on a much longer timescale.
In this Letter, we adopt a system consisting of several  
coupled pendula \cite{KK,Antoni_Ruffo}, each of which possesses two modes
of motion, oscillation and rotation. 
We clarify the characteristic dynamics necessary for energy storage
in connection with the coexistence of these two modes, and
also show that partial contact with the heat bath is necessary 
for this storage.
The relationship between  the storage time and the injected energy is
obtained, and is experimentally verifiable.
The relevance of our results to protein motors is also discussed.

A protein consists of a folded chain of amino acids that assumes a 
globular shape \cite{Protein_structure}.
The main chain is accompanied by side chains arranged around it,
with each side chain hanging on the main chain similar to a pendulum.
Some side chains are gathered in a globular shape,
and a certain assembly of them can play an important role in the
function of the protein.
As an abstract model for the angular motion of
side chains in such a functional assembly, 
we choose a system of coupled pendula.
In particular, we study the idealized case of $N$ identical pendula 
equally coupled to each other.  
Here, the oscillation of the pendula corresponds to the vibration of 
the side chains.
With this simple model as an example, we demonstrate that 
long-term storage is generally possible in a class of Hamiltonian systems
\cite{generality}.
Our study is restricted to a classical mechanical description, since
quantum mechanical effects are believed to be irrelevant to protein dynamics
(except for the choice of the potential). 

Our Hamiltonian is given by
\begin{equation}
H=K+V=\sum\limits_{i=1}^N {p_i^2\over 2}+\sum_{i,j=1}^N V(\theta_i,\theta_j),
\end{equation}
where $p_i$ is the momentum of the $i$-th pendulum.
The potential $V$ is constructed so that 
each pair of pendula interacts through their phase difference with an
attractive force to align the phases \cite{KK,Antoni_Ruffo}:
\begin{equation}
V(\theta_i,\theta_j)  =  
{1 \over {2(2\pi)^2N}}\{1-\cos(2\pi(\theta_i-\theta_j))\}.
\end{equation}
Hence, the evolution equations for the momentum $p_i$ and the phase 
$\theta_i$ are given by
\begin{eqnarray}
\dot p_j &=& {1\over{2\pi N}}\sum_{i=1}^N \sin(2\pi(\theta_i-\theta_j)),
\label{eqn:seijunP}\\
\dot \theta_j &=& p_j.
\label{eqn:seijunQ}
\end{eqnarray}
In this model, all pairs of pendula interact 
identically.  
If the protein were a straight chain without folding, 
it could be modeled by a one-dimensional chain of pendula.
In reality, due to its globular shape, a global interaction occurs
between pendula. Although the assumed ``mean-field coupling" with uniform
strength represents an extreme simplification, the resultant model can 
capture some general characteristics of protein dynamics.

For the (dynamical) function of proteins, it is essential that they exist
in aqueous solution, which functions as a heat bath.
Since the hydrophobic part of a protein molecule is
segregated from the aqueous solution, contact with
the heat bath is restricted.
Accordingly, we define the equation of motion so that only some pendula 
in the system contact the heat bath:
\begin{equation}
\dot p_i = -{\partial H \over {\partial \theta_i}} -\gamma p_i
+\sqrt{2\gamma T}\xi_i(t) \qquad (i\leq N_h).
\label{eqn:Langevin}
\end{equation}
Here, the heat bath is described by the Langevin equation,
in which $T$ represents its temperature,
$\gamma$ is a relaxation coefficient, and
$\xi_i(t)$ is a Gaussian random form satisfying
$\langle \xi_i(t)\rangle =0$ 
and $\langle \xi_i(t_1)\xi_j(t_2) \rangle
=\delta_{ij}\delta(t_2-t_1)$, with $\langle \cdot \rangle$ 
as the temporal average.
An important point here is that contact with the heat bath is 
restricted to the few pendula satisfying $i\leq N_h$.
This restriction may be interpreted
as only allowing for interaction of the hydrophilic part
with the aqueous solution.
Note that this restricted heat bath (i.e., partial contact with the heat bath)
is sufficient for realizing thermal equilibrium.

Here, we briefly review the typical behavior of 
eqs. (\ref{eqn:seijunP}) and ({\ref{eqn:seijunQ}) in the conservative case 
\cite{KK,Antoni_Ruffo,Yamaguchi,Latora_Ruffo}, without coupling 
to the heat bath.
When the total energy is small,
all pendula vibrate almost synchronously
(i.e., $|\theta_i-\theta_j|$ and $|p_i-p_j|$ are small for all $i$, $j$).
The typical timescale of vibration is $\sim 1$ - $10$,
while there is collective motion giving rise to
a periodic change in the degree of synchronization, with a
timescale of $\sim 10^2$ - $10^3$.

As the total energy increases, rotational (phase-slip) 
behavior of pendula begins to appear.
For a sufficiently large total energy,
almost all pendula exhibit rotational motion.
Here, each $p_i$ changes slowly over time, with relatively large 
values of $|p_i-p_j|$.
In this case, the pendula rotate almost freely, 
since the correlation of their phase with those of other pendula
cannot be maintained, and the interaction term cancels out even with 
a short time average,
except in the rare situation that two momenta take very close values.
The timescale of rotation is $O(1)$.

In the medium energy regime,
the rotational motion of a single pendulum appears intermittently 
from an assembly of vibrating pendula
(as displayed in the upper part of Fig. \ref{fig:eng_flc}).
Once a pendulum starts to rotate, it typically
continues to rotate over many cycles,
which is longer than the typical timescale of each pendulum's 
vibration (and rotation).
We emphasize here that 
the effective interaction for a rotating 
pendulum is {\it much weaker} than those for vibrating ones,
as mentioned above.

In the thermodynamic limit with $N \rightarrow\infty$ 
\cite{Antoni_Ruffo,Yamaguchi,Latora_Ruffo},
there appears a phase transition at the energy $E_c/N=0.0190$ ($T_c=0.0127$), 
where the vibrational mode is dominant in the `solid'-like phase and
the free rotation is dominant in the `gas'-like phase.
If $N$ is not sufficiently large (as is the case studied here), 
the transition point is blurred by the finite size effect, 
and the fluctuations are large around the transition point.
Hereafter, we refer to this temperature range 
($0.01 \stackrel{<}{\sim} T  \stackrel{<}{\sim} 0.03$)
as the {\it medium temperature} range.

Now, we consider the Hamiltonian system 
in contact with the heat bath described by eq. (\ref{eqn:Langevin}), where
the total energy is now time-dependent.
It is found that the fluctuations of energy are sometimes correlated with 
the dynamics of the pendula.
For instance, in the medium temperature regime in partial contact 
with the heat bath, the intermittent 
appearance of rotating pendula, similar to the Hamiltonian dynamics, 
becomes accompanied with a large fluctuation of the total energy
(see Fig. \ref{fig:eng_flc}).
Here, deviation toward
a larger total energy is supported by the concentration of energy
in one (or few) pendulum \cite{Takeno_Peyard}.

While local fluctuations are thus dependent on the dynamics,
the system reaches {\it thermal equilibrium} for a sufficiently 
long timescale,
irrespective of the values of $\gamma$, $N_h$ ($\geq 1$), or the temperature.
Note that the equilibrium property as a canonical ensemble is identical 
for any number of $N_h$, even for $N_h \ll N$.
When $N_h=N$ (full contact), 
the timescale to reach thermal equilibrium is determined completely by 
the value of $\gamma$,
independently of the dynamical properties of the system.
On the other hand, in the case of partial contact, 
the relaxation process is affected by inherent Hamiltonian dynamics.
It progresses beyond the timescale $\Gamma^{-1}$, 
where $\Gamma\equiv{\gamma N_h}/N$ gives the dissipation rate of the system.
Even in this case,
the dynamics of the pendula corresponding to $i\leq N_h$ are governed by
the timescale $\gamma^{-1}$ ($< \Gamma^{-1}$).

As the next step, we discuss the energetic behavior of the system 
{\it far from equilibrium}.
Consider a special enzymatic event such as ATP attachment or its reaction.
With such an event, proteins are moved far from equilibrium.
We study how such an event is related to the storage of energy.

Due to such a `reaction' event,
some portion of the protein is
forced far from the previous equilibrium state.
This situation is modeled by the addition of an instantaneous kick
to a certain pendulum at the reaction event.
The kicked pendulum comes to possess a larger amount of momentum 
and kinetic energy than other pendula 
\cite{absorption}.

An example of temporal evolution after the kick at $t=t_0$ is 
shown in Fig. \ref{fig:eng_str}, where the system with $N=10$ and $N_h=1$
is adopted; that is only the first pendulum is in contact with the heat bath.
Kinetic energy with the amount of $E_0$ is added 
to the kicked pendulum.
In Fig. \ref{fig:eng_str}, the high-energy state continues up to
$2\times 10^{5}$ while the typical relaxation time $\Gamma^{-1}$ 
of the heat bath is equal to $10^{3}$.
There, the kicked pendulum continues to rotate in isolation, maintaining
the large energy, while it is affected only slightly by the other pendula 
over a long time interval.
This allows for long-term energy storage.  

In order to study the lifetime of energy storage,
we define it as the interval from the kick until the relaxation of 
the total energy (see Fig. \ref{fig:eng_str}). This is determined
by the time $t_R$ at which the total kinetic energy $K$
decreases to $NT/2$, that is the value at thermal equilibrium.
The results, however, do not depend on the specific choice of
the relaxation time. 

The distribution of the lifetime is shown in Fig. \ref{fig:dstrb},
where evolutions for 200 stochastic processes $\xi_i(t)$ are sampled.
For comparison, we show the distribution for two cases
with the same value of $\Gamma$: $N_h=1$ with $\gamma=10^{-2}$ 
and $N_h=N(=10)$ with $\gamma=10^{-3}$.
It is noted that the distribution of the lifetime is quite different 
for the two cases.
For the case of partial contact, $N_h=1$,
the typical lifetime of the energy storage is very long 
and reaches $10^6$, in contrast with
$10^3$ for the  $N_h=N$ case. 
Moreover, the long-term energy storage is obtained 
as far as the kicked pendulum is not in contact with the heat bath,
even if $N_h\simeq N$ as long as $N_h \neq N$.
This result suggests that {\it partial contact with the heat bath 
is necessary for long-term energy storage.}

Although the thermal equilibrium properties are similar in the above two cases
($N_h < N$ and $N_h=N$),
a large difference appears when the system is placed far from equilibrium.
The pendula $i\le N_h$, interacting directly with 
the heat bath, cannot rotate freely over the $1/\gamma$ timescale
(in contrast with the Hamiltonian dynamics)
and the dynamics is replaced by Brownian motion due to the heat bath.
In the case of partial contact, 
a long duration of rotation is possible for pendula corresponding to $i>N_h$,
if the pendula possess sufficiently large energy
to remain far from equilibrium.
The pendula there become free from the thermal effect, due to
the distinctively weaker interaction with the other pendula,
and follow nearly pure Hamiltonian dynamics.
Although the prototype of this
mechanism is observed as a large fluctuation around equilibrium
in the medium temperature regime (see Fig. \ref{fig:eng_flc}),
it works well far from equilibrium.

The lifetime of energy storage increases with the increase of the
kicked energy $E_0$, injected to a single pendulum.
Figure \ref{fig:E0_lifetime} shows the relationship between the average 
lifetime and $E_0$.
The cases in partial contact $N_h=1$ with two different $\gamma$ values
are compared with the Hamiltonian (i.e., microcanonical) case 
without the heat bath \cite{note}.
In the Hamiltonian case, we find the following relation
\begin{equation}
\langle t_R-t_0 \rangle \propto \exp(\alpha \sqrt{E_0}),
\label{eqn:exponential}
\end{equation}
between the average lifetime and $E_0$ 
with $\langle\cdot\rangle$ as the ensemble average, and $\alpha$ 
as a temperature-dependent constant.
In a dissipative case with the heat bath,
a similar increase of the relaxation time is maintained,
although there exists slight suppression of the lifetime
at a high energy.

In Hamiltonian systems, the exponential law of relaxation time
is known as the `Boltzmann-Jeans conjecture'
\cite{Boltzman_Jeans1,Boltzman_Jeans2,Boltzman_Jeans3,Boltzman_Jeans4}.
The energy relaxation between slow and fast modes generally requires
a considerable amount of time. 
The conjecture is confirmed in numerical experiments for
a classical gas of diatomic molecules \cite{Benettin}, and
some analytic estimates for a generalized version of the conjecture
are given within a classical perturbation theory of
Hamiltonian systems\cite{Nekhoroshev}.
In our case, there is only slight energy exchange between the fast
rotational and slow vibrational modes.
In the above diatomic molecules with translation,
rotation, and vibration, 
when it is sufficiently high, the rotational energy almost freezes, 
and the transfer of the energy to the translational
mode requires a time exponential to the angular momentum,
similar to eq. (\ref{eqn:exponential}).
We may expect that this exponential form is generally valid,
if the excited mode has a weaker interaction with other modes
as the excited energy becomes higher.

Thus far, the exponential law has been proven for a class of 
Hamiltonian systems.
Although our model has partial contact with the heat bath, 
a similar increase of the relaxation time is confirmed.
Moreover, the same increase is obtained for a system with
the same $\Gamma$ value, for any value of $N_h$ (up to $N-1$),
as long as the rotating pendulum does not come into direct contact
with the heat bath.
On the other hand, if all pendula are in contact with the heat bath,
the energy of the rotating pendulum decays much faster than
the case with partial contact depicted in Fig. \ref{fig:E0_lifetime}.
The typical lifetime for this full contact case is of the order $\Gamma^{-1}$
over a wide range of $E_0$.
The system in partial contact with the heat bath retains a mechanism for
slow relaxation, similar to the Hamiltonian system.

What type of Hamiltonian is required for long-term energy storage?
Consider a system with two possible phases (e.g., gas-like and 
solid-like phases). 
Here, some elements possessing higher energy ``melt" 
and come to have a distinctively weaker interaction with elements 
in the `solid'-like phase.
Such differentiation is indispensable to the
storage of a large amount of energy.  
Partial contact with a heat bath is essential 
to maintain the differentiation of states required to store 
the absorbed energy.
Our coupled pendulum model provides a simple example for this behavior.

The dynamic mechanism for energy storage presented here
is simple enough to be realized in real protein motors.
It is only necessary that the protein dynamics  
exists near a phase transition region with `solid'-like (vibrational) modes
and `gas'-like (rotational) modes
and that the folded structure prevents some parts of the protein 
from experiencing the random effect of a heat bath.
These conditions are expected to be satisfied for many kinds of proteins 
with a sufficiently large size, in addition to the protein motors.
It is important to note that one can confirm experimentally if the present
mechanism of the energy storage is valid or not by examining the relationship
between the storage time and the injected energy, as shown
in Fig. \ref{fig:E0_lifetime}.

The authors are grateful to
T. Yanagida, Y. Ishii and members of the Single Molecule Processes Project, JST
for many meaningful suggestions.
They would also like to thank F. Oosawa,
T. Yomo, T. S. Komatsu, S. Sasa and T. Shibata
for stimulating discussions.  
This research was supported by Grants-in-Aids for Scientific 
Research from the Ministry of Education, Science, Sports and Culture of Japan.

\begin{figure}[h]
\caption{Large deviation of the total energy $E=K+V$ at thermal equilibrium,
accompanied by the occasional appearance of the rotational mode
(see the region $50000<$time$<60000$). $T=0.02$, $N=10$, $N_h=1$ 
and $\gamma=10^{-2}$.
In the upper figure, all time series of the momenta $p_i$ 
($1\le i \le N$) are overlaid.
}
\label{fig:eng_flc}
\end{figure}

\begin{figure}[h]
\caption{Process of energy storage after an instantaneous kick at $t=t_0$.
$T=0.02$, $\gamma=10^{-2}$, $N_h=1$, $N=10$ and $E_0=0.35$.
upper: Time series of the momenta $p_i$ ($1\le i \le N$). 
The kicked pendulum A relaxes to thermal equilibrium around $t_{R}$.
lower: The time series of the total energy $E$. The dotted line indicates 
the value at thermal equilibrium.
}
\label{fig:eng_str}
\end{figure}

\begin{figure}[h]
\caption{
Distributions of the lifetime $\log_{10}(t_R-t_{0})$
for energy storage at $T=0.02$. $N=10$ and $E_0=0.5$.
Each distribution is obtained from $200$ samples.
Unshaded distribution: $N_h=1$ with $\gamma=10^{-2}$.
Shaded distribution: $N_h=10$ with $\gamma=10^{-3}$.
}
\label{fig:dstrb}
\end{figure}

\begin{figure}[h]
\caption{
Average lifetime $\langle t_R-t_{0}\rangle$ versus the
square root of injected energy
$E_0$ for the system with $N=10$.
The ensemble average is computed from $100$ samples of different
initial conditions.
$\circ$ : Hamiltonian system with $2\langle K\rangle \simeq 0.02$.
$\times$ : dissipative system with $N_h=1$, $\gamma=10^{-3}$ and $T=0.02$.
$\triangle$ : dissipative system with $N_h=1$, $\gamma=10^{-2}$ and $T=0.02$.
The data from $N_h=9$ with the same $\Gamma$ value agree with
the case $N_h=1$, within the statistical error.
}
\label{fig:E0_lifetime}
\end{figure}


\begin{thebibliography}{999}
\bibitem{generality}
Instead of side chains,
the pendula can be regarded to represent
dipoles or peptide bonds, as a model of protein.
\bibitem{absorption}
The dynamical process of energy absorption into one pendulum, 
distinctively from an instantaneous kick,
is discussed in 
N. Nakagawa and K. Kaneko,
chao-dyn/9903005
\bibitem{note}
The relaxation time for conservative systems is defined as
the relaxation of the momentum $p_N$ for kicked pendulum.
\end{thebibliography}
\end{document}